# Feasibility of Energy Neutral Wildlife Tracking using Multi-Source Energy Harvesting

Samer Nasser
Samer.Nasser@uantwerpen.be
IDLab, University of Antwerp – imec
Antwerp, Belgium

Henrique Duarte Moura
Henrique.DuarteMoura@uantwerpen.be
IDLab, University of Antwerp – imec
Antwerp, Belgium

Dragan Subotic
Dragan.Subotic@uantwerpen.be
IDLab, University of Antwerp – imec
Antwerp, Belgium

Ritesh Kumar Singh
RiteshKumar.Singh@uantwerpen.be
IDLab, University of Antwerp – imec
Antwerp, Belgium

Maarten Weyn
Maarten.Weyn@uantwerpen.be
IDLab, University of Antwerp – imec
Antwerp, Belgium

Jeroen Famaey
Jeroen.Famaey@uantwerpen.be
IDLab, University of Antwerp – imec
Antwerp, Belgium

## ABSTRACT

Long-term wildlife tracking is crucial for biodiversity monitoring, but energy limitations pose challenges, especially for animal tags, where replacing batteries is impractical and stressful for the animal due to the need to locate, possibly sedate, and handle it. Energy harvesting offers a sustainable alternative, yet most existing systems rely on a single energy source and infrastructure-limited communication technologies. This paper presents an energy-neutral system that combines solar and kinetic energy harvesting to enable the tracking and monitoring of wild animals. Harvesting from multiple sources increases the total available energy. Uniquely, the kinetic harvester also serves as a motion proxy by sampling harvested current, enabling activity monitoring without dedicated sensors. Our approach also ensures compatibility with existing cellular infrastructure, using Narrowband Internet of Things (NB-IoT). We present a simulation framework that models energy harvesting, storage, and consumption at the component level. An energy-aware scheduler coordinates task execution based on real-time energy availability. We evaluate performance under realistically varying conditions, comparing task frequencies and capacitor sizes. Results show that our approach maintains energy-neutral operation while significantly increasing data yield and reliability compared to single-source systems, with the ability to consistently sample GPS location data and kinetic harvesting data every two minutes while transmitting these results over NB-IoT every hour. These findings demonstrate the potential for maintenance-free, environmentally friendly tracking in remote habitats, enabling more effective and scalable wildlife monitoring.

## 1 INTRODUCTION

Wildlife is essential to ecosystem health and balance. Shifts in animal populations often signal broader environmental issues such as pollution, climate change, or habitat degradation. By monitoring wildlife, scientists can track biodiversity changes and assess ecosystem resilience [2]. Movement tracking, in particular, reveals how species respond to human activities and environmental pressures, offering insights that are vital for global conservation strategies [19]. Furthermore, wildlife studies enhance our understanding of biology, behavior, and evolution, and data from tracking efforts informs conservation strategies, policy decisions, and innovation.

While animal tracking has a long history, the emergence of the Internet of Things (IoT) has significantly enhanced its capabilities. However, long-term tracking remains a challenge due to the energy limitations of non-rechargeable batteries [6]. In remote or rugged environments, tracking systems often need to operate autonomously for months or even years without maintenance. This makes energy autonomy a critical design requirement. Modern tracking devices typically integrate power-hungry components such as GPS, wireless communication modules, and environmental sensors, further increasing energy demands. These challenges are even more pronounced in small or lightweight animal tags, where strict size and weight constraints limit battery capacity. Environmental concerns also arise from the widespread use of disposable batteries, which contribute to electronic waste and introduce toxic materials into ecosystems [25]. These factors have driven a growing interest in energy-harvesting technologies that draw power from ambient sources such as solar radiation [23], kinetic motion [5, 24], fluid flow [12, 14], and thermal gradients [1]. When combined with energy storage devices like rechargeable batteries or supercapacitors, these harvesting methods enable more sustainable and longer-lasting tracking solutions. In our work, we opt for using supercapacitors because of their longer lifetime, absence of toxic and rare earth materials, and greater resilience to extreme temperatures as compared to rechargeable batteries, making them especially well-suited for long-term wildlife tracking in harsh or inaccessible environments.

Bäumker et al. [1] developed a wildlife tracking system powered by thermal energy harvested from an animal's body heat. Their device enabled GPS position fixes every 1.1–1.5 hours, collected temperature data every 7 minutes, and transmitted data via LoRaWAN every 14 minutes. Gregersen et al. [5] proposed a similar system using kinetic energy harvested from an animal's motion. In trials with domestic dogs, it generated 2.26–3.2 joules daily, with an average of 1.48 hours of walking. Experiments with an Exmoor pony yielded only 0.69 joules daily, limiting GPS fix and Sigfox transmissions to once every few days. The transmissions included accelerometer data sampled once per day.

Both systems demonstrate the potential of capacitor-based tracking but rely on single energy sources and communication infrastructures that may be unavailable in remote areas. To address these limitations, our work proposes a multi-source energy harvesting system combining solar and kinetic energy and utilizing NB-IoT for connectivity. This approach improves energy efficiency and availability, and enables broader scalability using existing cellular networks. Additionally, the kinetic harvester is leveraged as a proxy for activity monitoring by sampling harvested current, allowing for movement intensity inference and, effectively, a dual-function harvester-sensor.

To evaluate the feasibility and performance of this multi-source, capacitor-based tracking system, this paper presents a simulation framework that models energy harvesting, storage, and consumption at the component level. We introduce energy-aware scheduling to coordinate tasks based on available energy, and we compare different task execution frequencies and capacitor sizes to assess their impact on system reliability, data yield, and overall energy neutrality under realistic environmental conditions and hardware requirements.

The remainder of this article is organized as follows: Section 2 describes the system architecture and components used in this work. An analysis of the energy consumption of those components in their different states is described in Section 3. Section 4 contains information about the capacitor behavior and the system simulation. Section 5 describes the methodology and evaluates the system for multiple scenarios. Finally, Section 6 discusses the conclusions, challenges, and future outlooks.

## 2 SYSTEM ARCHITECTURE

The proposed system enables energy-neutral wildlife tracking using a supercapacitor for storage, ensuring ultra-low power consumption. It integrates multi-source ambient energy harvesting, efficient data processing, and low-power GPS localization and wireless communication to minimize energy use. The system must collect GPS data and perform additional sensing multiple times per day to gather behavioral insights. An overview of the system architecture can be found in Figure 1. It consists of three main components that work together to optimize energy use: (1) Energy Harvesting and Storage, (2) Energy Management and Controls, and (3) Peripherals. The final system integrates energy-aware scheduling mechanisms, ensuring continuous operation without depleting power.

### 2.1 Energy Harvesting and Storage

To power our GPS wildlife tracking system, we employ a dual energy harvesting approach that combines solar and kinetic sources. Solar energy harvesting serves as the primary source due to its high yield under favorable lighting conditions. However, its major limitation is the inability to generate energy during nighttime or extended low-light periods. To address this, we integrate a kinetic energy harvester that continues to produce energy when solar input is unavailable, particularly at night. Beyond supplementing energy supply, the kinetic harvester also contributes to behavioral sensing. By monitoring the current generated by the kinetic harvester over time, we obtain a proxy for the animal's movement intensity, enhancing the sensing capabilities of the system without incurring significant overhead.

Solar energy is harvested through the use of a small solar panel connected to a Power Management Integrated Circuit (PMIC) [17] which leverages Maximum Power Point Tracking (MPPT) to maximize energy extraction and efficiency. Kinetic energy is extracted using a gravity-based generator that converts motion into electrical energy through a swinging ferromagnetic pendulum [5, 8]. The resulting kinetic energy is also routed to a PMIC [15], which features a lossless Coulomb counter that continuously monitors the incoming current over time. Because this current is proportional to the intensity of the animal's movement, it can be used to infer behavioral patterns, for instance, detecting periods of localized movement even when the GPS position remains relatively unchanged or when there is no GPS data available. This data can also be used to detect if the animal has been standing still, moving, or running. Thus, the PMIC fulfills a dual role as both a power management element and an indirect movement sensor.

Both PMICs offload the harvested energy into a custom-developed board, described in detail in [10], which efficiently combines the two input sources and prevents unintended energy feedback loops. These loops occur when energy flows unintentionally between sources or converters, leading to inefficiency and potential instability. The combiner prevents this by using separate temporary capacitor buffers for the different harvesting sources, which then transfer their energy to a shared supercapacitor in a coordinated manner [10].

### 2.2 Energy Management and Controls

The supercapacitor supplies energy to the Microcontroller Unit (MCU), the system's main controller. Unlike batteries, capacitors store energy electrostatically rather than chemically, enabling rapid charge and discharge cycles. This makes them ideal for intermittent energy harvesting, but their limited energy density and voltage-dependent nature require careful energy management. To prevent depletion—defined as the capacitor voltage dropping below a critical threshold $V_{min}$—the MCU employs an energy-aware task scheduler (described in Section 2.4). This critical threshold is defined by the minimal required input voltage for the buck-boost converter, which up- or down-converts the fluctuating voltage from the supercapacitor to a stable supply voltage. If depletion occurs, potentially causing data gaps, the MCU remains inactive until the capacitor recharges to a predefined turn-on voltage $V_{\mathrm{to}}$. Moreover, if depletion occurs during task execution, then its outcome is lost, wasting energy. To mitigate this, preconfigured voltage thresholds are set for each task based on prior power consumption measurements. These conservative thresholds ensure safe execution even in the absence of ongoing energy harvesting. The system prioritizes stability by executing tasks only when sufficient charge is available, preventing failures and optimizing energy use.

### 2.3 Peripherals

The system regularly performs GPS fixes to determine the animal's location. The GPS module communicates the location data to the MCU via the $I^2C$ interface. The MCU then stores this time-stamped data in internal flash memory. To reduce communication overhead, the MCU is then configured to transmit the stored location data—along with the Coulomb counter readings described in Section 2.1—to the cloud in bulk every hour using the NB-IoT module.

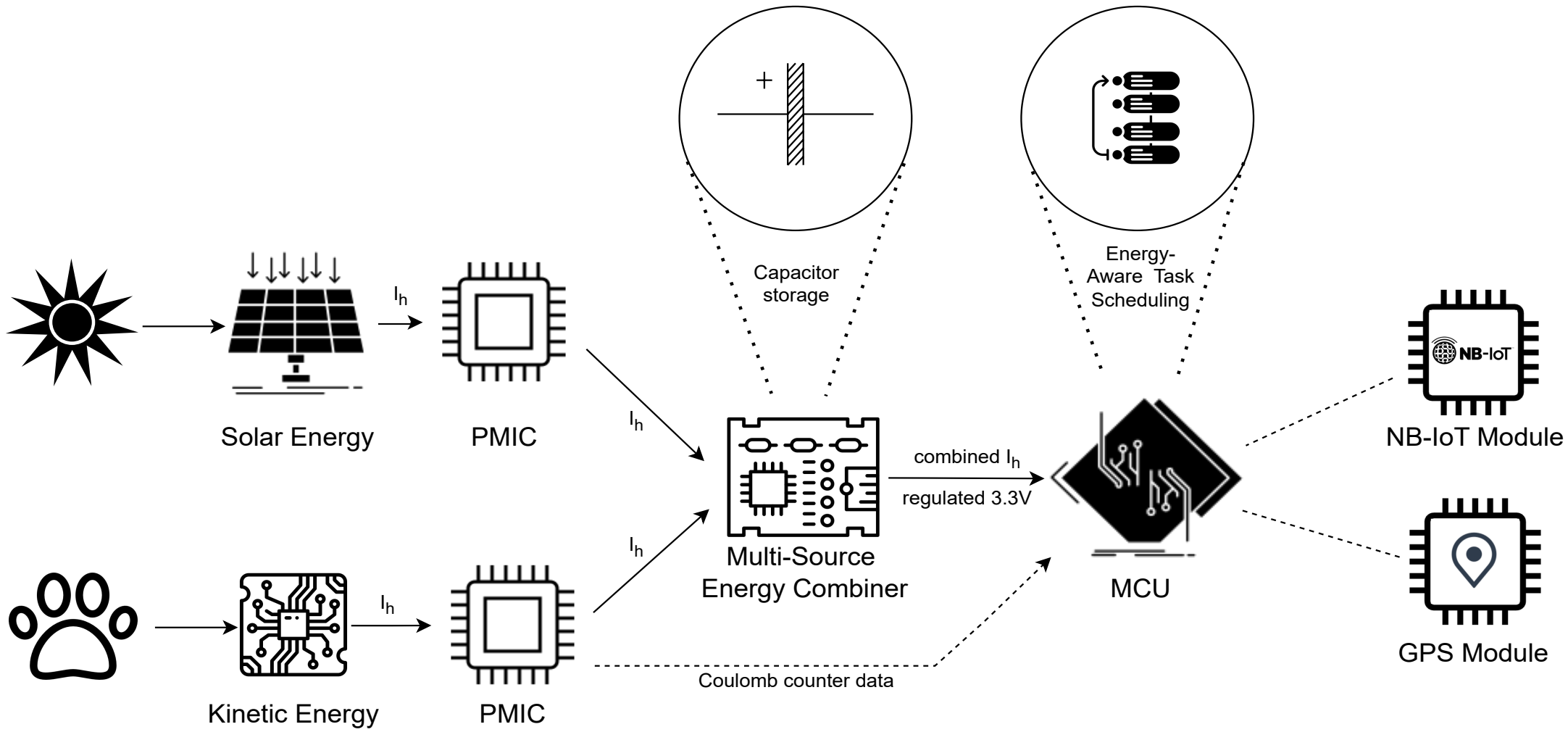

**Figure 1: Overview of the system architecture of the proposed energy-neutral wildlife tracker. Ambient energy from solar and kinetic harvesters gets optimally extracted through their respective PMIC, to then be combined using a custom multi-source energy combiner containing a supercapacitor as storage element. From there the voltage gets regulated to a stable 3.3 V to power the MCU, which runs an energy aware task scheduling algorithm to efficiently perform GPS localization and NB-IoT communication. Furthermore, kinetic harvesting data is captured through the use of a lossless Coulomb counter on the PMIC to provide auxiliary data on the animal's intensity of movement.**

To enable these functionalities, two peripheral modules are interfaced with the MCU. An NB-IoT module provides robust wireless communication capabilities. As a Low-Power Wide-Area Network (LPWAN) technology, NB-IoT is specifically designed for efficient communication between low-power devices and cellular networks. Operating on licensed frequency bands, it offers extensive coverage, low energy consumption, and high scalability, making it particularly well-suited for large-scale deployments. Its widespread global adoption ensures compatibility and reliable connectivity across various geographic regions, which is advantageous for this application.

For geolocation, a low-power Global Navigation Satellite System (GNSS) module is integrated to determine the outdoor position of the animal. We consider it to be operating in GPS-only mode for power efficiency and refer to it as the GPS module for the remainder of this paper. The GPS module supports three primary operating modes: *hot start*, *warm start*, and *cold start*. These modes differ in terms of their time to fix and energy consumption, and are selected based on how recently the ephemeris data was refreshed. Ephemeris data consists of satellite orbit parameters broadcast by the satellites, which are required for *hot start*.

The system defaults to *hot start* mode, which offers the fastest fix with the lowest energy cost, provided the ephemeris data is no older than four hours. If the data becomes outdated, the system switches to *warm start*, which leverages AssistNow Autonomous [21] to reconstruct satellite orbits based on outdated ephemeris data to perform localization. *Warm start* requires significantly more energy—approximately four times that of a *hot start*—making timely ephemeris updates essential. In our implementation, the GPS module performs an ephemeris update by remaining powered on for a short period following a fix, and downloading the updated data directly from the satellites. If the ephemeris data is not refreshed for more than two days, or the system has experienced a power failure, the GPS module defaults to a *cold start*. This is the most energy- and time-intensive mode, which operates without any prior localization or orbit information.

To reduce the system's overall power consumption, the connections between the MCU and the peripheral modules are managed using MOSFETs. Specifically, the main power line to each peripheral module can be completely disconnected when the peripheral is idle. However, in the case of the GPS module, the backup power line remains continuously connected. This ensures that the module can maintain its backup RAM and Real-time Clock (RTC), which are essential for supporting future *hot start* operations, while minimizing idle power draw.

The GPS scheduling strategy is implemented within the MCU's energy-aware task scheduler. Based on the freshness of the ephemeris data and the system's available energy, the appropriate GPS mode is selected to optimize energy efficiency.

## 2.4 Integrated System Behavior

Having described the key components, we now present the integrated behavior of the complete system. By coordinating energy-aware decision-making across all subsystems, the system achieves reliable operation while maintaining energy neutrality. The overall behavior is illustrated in Figure 2 and detailed in Algorithm 1. The system periodically cycles through sleep, sensing, and task execution phases, depending on current energy conditions and scheduled task requirements. The thick dashed arrows in Figure 2 indicate the start of an operation interval, where the system becomes active either from the *OFF* state or while already powered on. The duration of each interval is governed by how frequently the capacitor voltage $V_t$ is measured, which in turn determines when scheduling decisions are made.

At the start of each interval, the system enters the *SCHEDULE TASK* block, where it executes the scheduling logic defined in Algorithm 1. Here, we distinguish between two classes of tasks:

- **Main tasks**, such as GPS operations and NB-IoT communication, which are energy-intensive and subject to scheduling based on voltage thresholds ($V_{\text{thresh}}$), as discussed in Section 2.2.
- **Sensing tasks**, such as reading the capacitor voltage or the Coulomb counter, which are very low energy tasks executed without threshold constraints.

Analyzing Algorithm 1, we can see that by default, no main tasks are scheduled, and only the capacitor voltage is read as the default sensing task. For demonstration purposes, the system is configured to schedule a GPS fix every two minutes as a main task, along with a Coulomb counter reading as a sensing task. Depending on the freshness of the ephemeris data, the GPS fix is followed by downloading ephemeris data, and the fix happens in *hot_start* or *warm_start*.

Once every hour, the system schedules a high-priority NB-IoT transmission of the collected GPS and Coulomb data. The highest scheduling priority is reserved for a *cold_start*, which is only triggered if the ephemeris data is older than two days or if the system has experienced a full power loss. For each scheduled main task, the required voltage threshold $V_{\text{thresh}}$ is updated accordingly.

Following the scheduling step in Figure 2, the system proceeds to the *MEASURE* block to perform the sensing tasks. If the system remains powered and the measured voltage exceeds the threshold for the scheduled main tasks, it transitions to the *EXECUTE* block to carry out those tasks. A successful execution leads the system into the *SLEEP* state, where it remains idle until the end of the current interval. If a power failure occurs at any point in the interval, the system returns to the *OFF* state and awaits sufficient energy recovery to resume operation.

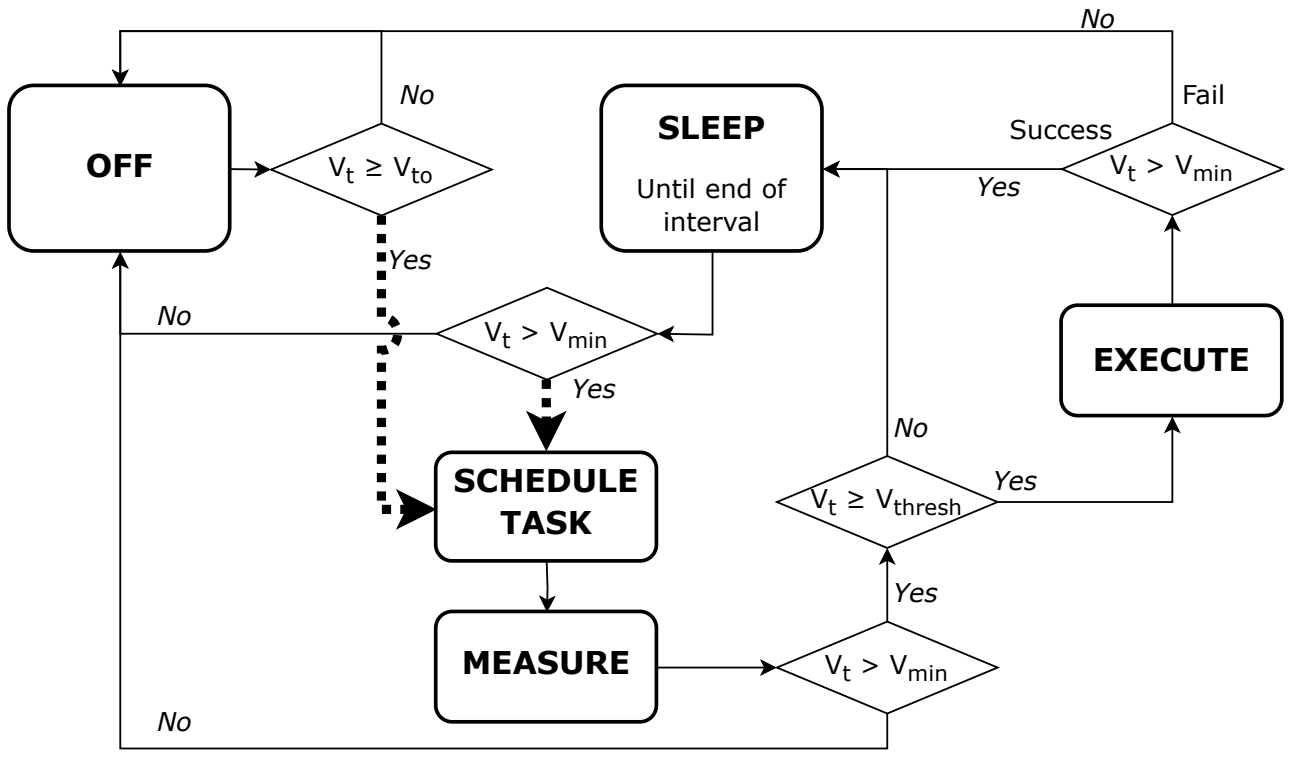

**Figure 2: Flowchart of the integrated system behavior, starting from the thick dashed lines. Tasks get scheduled according to Algorithm 1, followed by an execution of the ultra-low-power sensing tasks. Subsequently, the main tasks get executed if the capacitor is sufficiently charged. Finally, the system enters the sleep state until the start of the next interval.**

**Algorithm 1:** Task Scheduling Process

main_tasks = { Do nothing };
sensing_tasks = { Read cap voltage };
**if** *Full restart or ephemeris data is older than 2 days* **then**
  main_tasks = { cold_start };
  adjust $V_{thresh}$ accordingly;
**else if** *1 hour has passed* **then**
  main_tasks = { NBIoT };
  adjust $V_{thresh}$ accordingly;
**else if** *2 minutes have passed* **then**
  **if** *ephemeris data is older than 4 hours* **then**
    main_tasks = { warm_start, ephemeris download };
    adjust $V_{thresh}$ accordingly;
  **else if** *ephemeris data is older than 3 hours* **then**
    main_tasks = { hot_start, ephemeris download };
    adjust $V_{thresh}$ accordingly;
  **else**
    main_tasks = { hot_start };
    adjust $V_{thresh}$ accordingly;
**if** *2 minutes have passed* **then**
  sensing_tasks = { Read cap voltage, Read Coulomb };
**Schedule** main_tasks;
**Schedule** sensing_tasks;

## 3 ENERGY CONSUMPTION ANALYSIS

To ensure the long-term, maintenance-free operation of the tracking system, it is essential to understand how energy is consumed by its different components. This insight enables informed decisions about energy budgeting, task scheduling, and storage capacity, ensuring the system remains active across changing environmental conditions and usage patterns. Therefore, we analyze the power profile of the full system by breaking it down into subsystems, each with distinct energy requirements.

An overview is given in Table 1. The energy consumption is calculated considering a supply voltage of 3.3 V for the system since this is the minimal required supply voltage for the NB-IoT module. Furthermore, it should be stated that the values in each row of the table reflect the energy consumption contributions of separate modules, and hence, do not represent the system-level consumption. However, the system-level energy consumption can be derived from it and is further discussed in Section 3.5.

**Table 1: Energy Consumption Per Component Per Task ($V_{supply}$ = 3.3V)**

| Component | Current (mA) | Duration (s) | Energy (mJ) |
|---|---|---|---|
| GPS hot start | 7.5 | 1.0 | 24.75 |
| GPS warm start | 7.5 | 4.0 | 99.0 |
| GPS ephemeris download | 7.5 | 30.0 | 742.5 |
| GPS cold start | 8.0 | 36.118 ± *1.96* | 953.515 |
| GPS $I^2C$ write | 2.0 | 0.00038 | 0.0025 |
| GPS hardware backup | 0.028 | – | – |
| MCU Sleep (standby) | 0.00065 | – | – |
| NB-IoT | 20.65 ± *2.78* | 7.89 ± *1.66* | 537.664 |
| MCU ADC read | 0.311 | 0.00005 | 0.00005 |
| MCU $I^2C$ read Coulomb | 0.091 | 0.00023 | 0.00007 |
| MCU active base | 0.091 | – | – |
| Capacitor leakage | 0.03 | – | – |

## 3.1 GPS Localization

The first four rows of Table 1 describe the energy consumption of the GPS module in different operating modes: *Ephemeris download*, *hot start*, *warm start*, and *cold start*. The first three values are based on the SAM-M10Q module datasheet and integration manual [20, 21], as well as the GPS Interface Specifications [22]. For the *cold start* duration, the estimate is derived from measurements taken on the FireFlyX1 GPS module [4], which provides a representative time-to-fix estimate. In addition to localization, the system performs an Inter-Integrated Circuit ($I^2C$) write operation after each GPS fix to communicate location data to the MCU. This includes 8 bytes for longitude and latitude, and 4 bytes for GPS time, totaling 12 bytes (96 bits). At a bus speed of 400 kbps ($I^2C$ fast-mode), the estimated time for this 96-bit write is approximately 0.38 ms. This includes 0.33 ms for bus transfer (based on 132 bits at 2.5 $\mu$s per bit, including protocol overhead) and an estimated Central Processing Unit (CPU) processing overhead of 50 $\mu$s, assuming a 1 MHz CPU clock and polling mode. Finally, the energy consumption of the GPS module in *hardware backup* mode is considered. This ultra-low-power state maintains the module's RTC and backup RAM while consuming a minimal baseline current.

## 3.2 NB-IoT

The power profile for the NB-IoT modem was obtained through measurements on the nRF9160 NB-IoT module [9] using the Keysight N6705B precision power analyzer [7]. The process of waking up the radio module, registering to the network, reading 480 bytes of data from flash, and finally transmitting this data, was captured ten times in an outdoor environment. 480 bytes is a representative payload size since this corresponds to the amount of gathered data in one hour when reading the Coulomb counter data and GPS data every two minutes. The data in the fifth row of Table 1 shows the average and standard deviation of these measurements. The standard deviation is mainly attributed to the variability in the network attachment phase, particularly due to occasional preamble retransmissions during the random access procedure in NB-IoT.

## 3.3 Sleep

As the system is specifically optimized for energy efficiency, it spends most of its time in the ultra-low-power sleep state, where it tries to conserve energy in between activities. The current consumption described in Table 1 is based on the datasheet for the ultra-low power STM32L496 MCU [16]. While sleeping, the MCU is considered to be in standby mode with RTC functionality on, which translates to a consumption of 650 nA.

## 3.4 Sensing

The power profiles for the sensing segments are also built on the datasheet of the STM32L496. For the Analog-to-digital converter (ADC) read, representing the capacitor voltage measurement needed for energy-aware operation, we consider a sample time of 47.5 cycles with a CPU clock frequency of 1 MHz and a sample frequency of 1 Msps. Taking into account the ADC consumption of 220 $\mu$A and the MCU active current consumption of 0.91 mA, this results in a total estimated current of 311 $\mu$A for a duration of 50 $\mu$s. Similarly, the power profile for the $I^2C$ read, representing the Coulomb counter measurement, can be approximated. The current consumption is considered equal to the MCU active base current consumption. The $I^2C$ read duration for the Coulomb counter data can be estimated similar to the $I^2C$ write operation in Section 3.1, now considering 4 bytes of data and a slightly higher protocol overhead. At a bus speed of 400 kbps, the estimated time for a 32-bit $I^2C$ read operation is approximately 0.23 ms, including around 0.20 ms for the $I^2C$ bus transfer (based on approximately 80 bits at 2.5 $\mu$s per bit, including protocol overhead such as repeated start and direction change) and an estimated CPU processing overhead of about 30 $\mu$s, assuming 1 MHz CPU clock frequency and polling mode operation.

## 3.5 System-Level Energy Consumption

Building on the individual component-level consumption presented in Table 1, we now derive the system-level energy consumption per task. An overview is provided in Table 2. The first two columns indicate whether the active base current of the MCU (*MCUActiveBase*) and the GPS hardware backup current (*GPSBackup*) must be included in the total current draw for a given task.

For all peripheral tasks—such as GPS and NB-IoT communication—the MCU remains active as it functions as the system's central controller. Therefore, its base active current is added in those cases. Furthermore, for all tasks that do not involve active GPS operation, the GPS remains in its ultra-low-power backup mode, and its hardware backup current is included accordingly, as discussed in Section 3.1.

The current values listed in Table 2 also include the leakage current of the supercapacitor, as reported in the final row of Table 1. This parasitic loss arises from imperfections in the dielectric material, which allow a small continuous current to leak across the plates. The value is based on the datasheet of the DGH Series Supercapacitors [3], considering a 5.5 V, 5 F supercapacitor. As in Table 1, the energy consumption for each system-level task is calculated by multiplying the total current by the duration of the task and the supply voltage of 3.3 V.

**Table 2: System-Level Energy Consumption Per Task ($V_{\text{supply}}$ = 3.3 V)**

| Task | MCUActiveBase | GPSBackup | Current (mA) | Time (s) | Energy (mJ) |
|---|---|---|---|---|---|
| Hot start | ✓ | | 7.621 | 1.0 | 25.15 |
| Warm start | ✓ | | 7.621 | 4.0 | 100.6 |
| Eph. download | ✓ | | 7.621 | 30.0 | 754.5 |
| Cold start | ✓ | | 8.121 | 36.118 | 967.9 |
| GPS $I^2C$ write | ✓ | | 2.121 | 0.00038 | 0.00266 |
| Sleep | | ✓ | 0.05865 | – | – |
| NB-IoT | ✓ | ✓ | 20.799 | 7.89 | 541.5 |
| ADC read | | ✓ | 0.369 | 0.00005 | 0.000061 |
| $I^2C$ read Coul. | | ✓ | 0.149 | 0.00023 | 0.000113 |
| Turned off | | | 0.03 | – | – |

# 4 SYSTEM MODELING

To assess the energy neutrality and operational reliability of the proposed wildlife tracking system, we simulate its behavior over time under varying energy conditions and task schedules. Central to this simulation is the modeling of the supercapacitor, whose voltage governs the system's ability to execute tasks. Accurate modeling of the capacitor's charge and discharge dynamics is therefore essential to replicate real-world performance.

This section is divided into two parts. First, we present a mathematical model that describes the evolution of the capacitor voltage over time, capturing the effects of both energy harvesting and consumption. Then, we discuss the simulation environment and the conditions used to emulate realistic operating scenarios.

## 4.1 Capacitor Modeling

The charging and discharging behavior of a capacitor is dependent on the capacitance, the incoming harvested current, and the load it powers. The electrical model shown in Figure 3 illustrates the energy flow within the system, highlighting the key components responsible for harvesting, storing, and consuming energy. This model, based on the work in [13], represents the energy harvesting unit as a current source $I_H$, which captures power from environmental sources such as solar radiation, mechanical vibrations, or radio frequency signals. In our system, a multi-source energy combiner aggregates multiple inputs into a single equivalent current source.

At the core of the energy storage mechanism is the capacitor $C$, which accumulates energy over time. Its voltage, denoted $V_t$, evolves dynamically in response to the net current flow. The capacitor charges when the harvesting current exceeds the load demand, and discharges when the load draws more current than is harvested.

The load includes all energy-consuming elements of the system, such as the CPU, GPS, radio, and inherent capacitor leakage. These are collectively modeled as an equivalent resistance $R_{\text{eq}}$, which governs the rate of discharge.

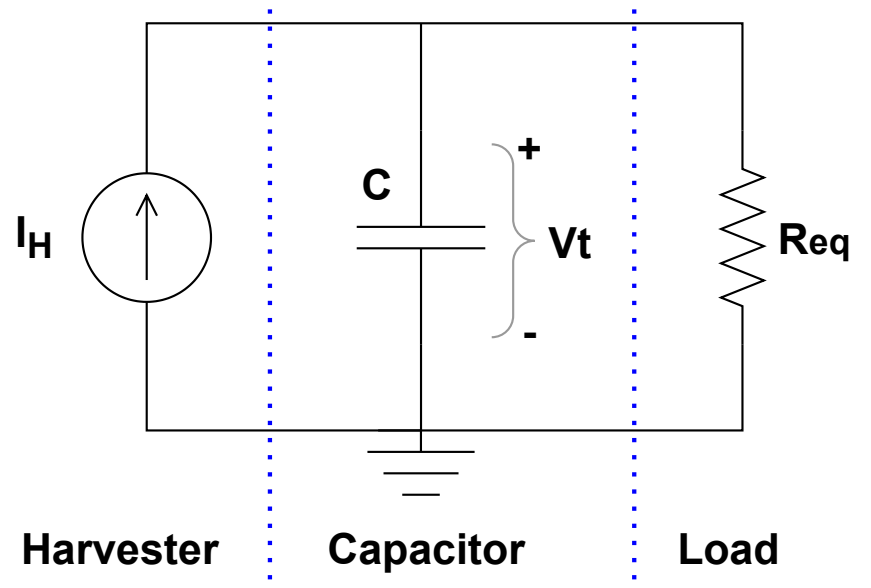

**Figure 3: Electrical model of the system architecture with three main components: Harvesting System, Energy Storage (supercapacitor), and Load.**

By taking these factors into account, we can approximate the capacitor voltage dynamics through mathematical modeling as described in Equation 1.

$$V_{\text{t+1}} = I_{\text{H}} \cdot R_{\text{eq}} \cdot \left(1 - e^{-\frac{\Delta t}{R_{\text{eq}} \cdot C}}\right) + V_t \cdot e^{-\frac{\Delta t}{R_{\text{eq}} \cdot C}} \quad (1)$$

The value of the capacitor voltage at time $t+1$, represented as $V_{\text{t+1}}$, is calculated here based on the value of the current (time $t$) capacitor voltage $V_{\text{t}}$, the incoming energy harvesting current $I_{\text{H}}$, the length of the time step $\Delta t$, the capacitance $C$, and the equivalent resistance for the power consumption of the system $R_{\text{eq}}$. $R_{\text{eq}}^{(a)}$ can be calculated for a specific task $a$ as $V_{\text{supply}}$ / $I^{(a)}$. For the supply voltage $V_{\text{supply}}$, we assume a fixed voltage of 3.3V as mentioned in Section 3. $I^{(a)}$ represents the current consumption of the system during the execution of task $a$. As discussed in Section 3.5, an overview of how these system-level current consumptions per task are obtained is given in Table 2.

## 4.2 Simulation Environment

Since this work is built on simulation, it is important to understand which data was used to represent the incoming solar and kinetic harvesting current.

For modeling the energy harvested from sunlight, we start from the solar irradiance. To this end, we used the private dataset from Puluckul and Weyn [11], which contains solar irradiance ($W/m^2$) measurements captured in Antwerp, Belgium, from November 24 to December 7 of 2023, with a time resolution of 1 minute. As this is data from wintertime, it represents worst-case performance. From this data, we can obtain an approximation of the incoming solar harvesting power. We assume a solar panel with dimensions of 40x40 mm and an efficiency of 18.5%. Furthermore, we take into account the cosine loss of the solar irradiance due to changes in the angle of incidence. Since this angle is dependent not only on the time of day but also on the movements of the animal, it is challenging to dynamically determine this value. Therefore, we use a conservative static approximation of 0.5 (60° angle with the vertical). Taking these values into account, we arrive at the approximated power generated by the solar panel with a conversion factor of 0.000148. The power can then be converted to harvested current by dividing it by the supply voltage of the MCU (3.3 V), which it is regulated to by the PMIC. Finally, based on the datasheet [17], we

**Table 3: Parameter Values Used in the System Model**

| Parameter | Value |
|---|---|
| Capacitor size | 1 F, 2.5 F, 5 F |
| Capacitor leakage | 10 µA, 16 µA, 30 µA |
| $V_{max_cap}$ | 5.5 V |
| $V_{supply}$ | 3.3 V |
| $V_{min}$ | 1.8 V |
| $V_{to}$ | 2.2 V |
| $V_{thresh_HS}$ | 1.9 V |
| $V_{thresh_HE}$ | 2.0 V |
| $V_{thresh_WE}$ | 2.1 V |
| $V_{thresh_NBIoT}$ | 2.0 V |
| Sense interval | 60 s |
| Fix interval | 60 s, 120 s, 300 s |

*Note: Capacitor sizes and leakage currents are paired in order.*

consider the PMIC to have an efficiency of 85%. This results in a solar conversion factor of 0.000038121.

The modeled kinetic harvesting current is based on the findings of Gregersen et al. [5], estimating wolves' average daily harvested energy at 13.07 J. Combining with the insights of the daily movement patterns of wolves for dawn, day, dusk, and night from Theuerkauf et al. [18], we can simulate a kinetic energy harvesting pattern for wolves using a 1-minute resolution, and convert harvested energy into current. The simulation assumes piecewise-constant energy harvesting between sample points, simplifying analysis while acknowledging real-life fluctuations. When combining solar and kinetic harvesting, a 0.88 efficiency factor [10] is applied to the multi-source combiner.

# 5 RESULTS AND DISCUSSION

This section shows the results of the simulated system behavior under varying configurations. We evaluate how different GPS fix intervals and capacitor sizes impact energy stability and tracking continuity. The analysis is based on the energy harvesting dynamics, voltage thresholds, and task scheduling constraints discussed earlier.

## 5.1 Experimental Parameters

Table 3 provides an overview of the parameters used in the experimental simulations. These values are based on the system architecture and energy considerations discussed in the previous sections. To evaluate the impact of different capacitor sizes, simulations were performed with three distinct capacitor values, each associated with a corresponding leakage current as shown in the first two rows of Table 3. All capacitors are rated for a maximum voltage ($V_{max_cap}$) of 5.5 V, hence the capacitor voltage never exceeds this value. A supply voltage of 3.3 V is used to calculate the equivalent resistance ($R_{eq}$) for each task as described in Section 4.1, based on Table 2. The minimum operating voltage $V_{min}$ is set to 1.8 V, corresponding to the minimal required input for the buck-boost converter on the energy combiner. A turn-on voltage $V_{to}$ of 2.2 V is used to maintain a safe buffer above $V_{min}$. Task-specific thresholds are defined based on energy requirements, with suffixes HS, HE, and WE in Table 3 referring to 'hot start', 'hot start + ephemeris download', and 'warm start + ephemeris download', respectively. These last values are experimentally determined, but not specifically optimized here. The capacitor voltage is sampled every 60 seconds, and simulations are run using GPS fix intervals of 1, 2, and 5 minutes to study the effect of task frequency on the capacitor voltage behavior.

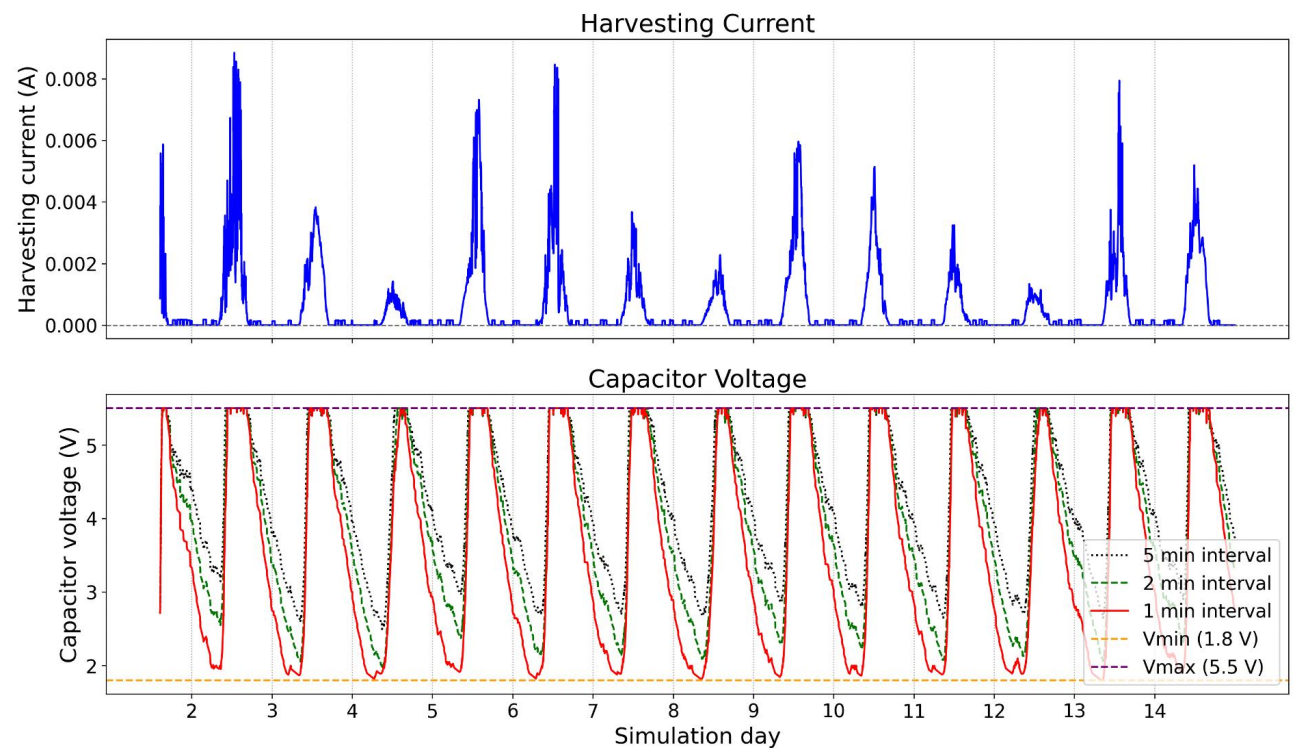

**Figure 4: Full simulation overview of the combined harvesting current and the capacitor voltage for a 2.5 F supercapacitor for different GPS fix intervals.**

## 5.2 Experimental Evaluation

This subsection presents the simulated performance of the system under various operational scenarios. We analyze capacitor voltage dynamics, energy availability, and GPS fix success under different GPS fix intervals and capacitor sizes.

*5.2.1 Full-Scale System Behavior.* Figure 4 presents the simulated behavior of the combined energy harvesting current and capacitor voltage over the full simulation period for three GPS fix intervals, using a 2.5 F supercapacitor. The evaluated fix intervals are 5 minutes, 2 minutes, and 1 minute—the latter matching the measurement frequency of the capacitor voltage and representing the maximum resolution in the simulation. A general inspection of Figure 4 reveals that solar energy harvesting dominates over kinetic harvesting. During daylight hours, the capacitor consistently charges to its full capacity across all three configurations. At night, when solar input is absent, the system relies solely on intermittent kinetic energy to sustain operation. In this phase, the impact of the fix interval becomes apparent: shorter intervals result in a steeper discharge curve, indicating higher energy consumption. A steeper curve implies the capacitor voltage approaches $V_{min}$ more quickly, reducing the available energy buffer during the night. Without adequate buffer sizing, the risk of energy depletion increases—particularly on days with lower-than-average energy yields compared to those represented in the dataset.

*5.2.2 Close-up Analysis.* Figure 5 presents a detailed view of the system's nighttime behavior between days 12 and 13. Variations in the voltage curve slope correspond to specific simulated events. Steep voltage drops indicate the execution of energy-intensive tasks, such as ephemeris downloads or NB-IoT transmissions. The effect of kinetic energy harvesting is highlighted between the dashed vertical lines. During such kinetically active periods, the harvested energy is sufficient to noticeably reduce the capacitor's discharge rate for the 1-minute fix interval, and even causes charging in the longer interval cases. The worst-case behavior for the 1-minute interval scenario begins at the square marker, where the capacitor voltage falls below the threshold required for a GPS hot-start. From this point onward, no GPS data is collected until the voltage recovers to 1.9 V. Further into the simulation, the voltage drops below $V_{\text{min}}$ (marked by the triangle), triggering a complete system shutdown. Recovery only occurs once the capacitor recharges to the turn-on voltage of 2.2 V, indicated by the circle. During this shutdown period, the system is unable to perform any tasks, resulting in a discontinuity in the tracking data.

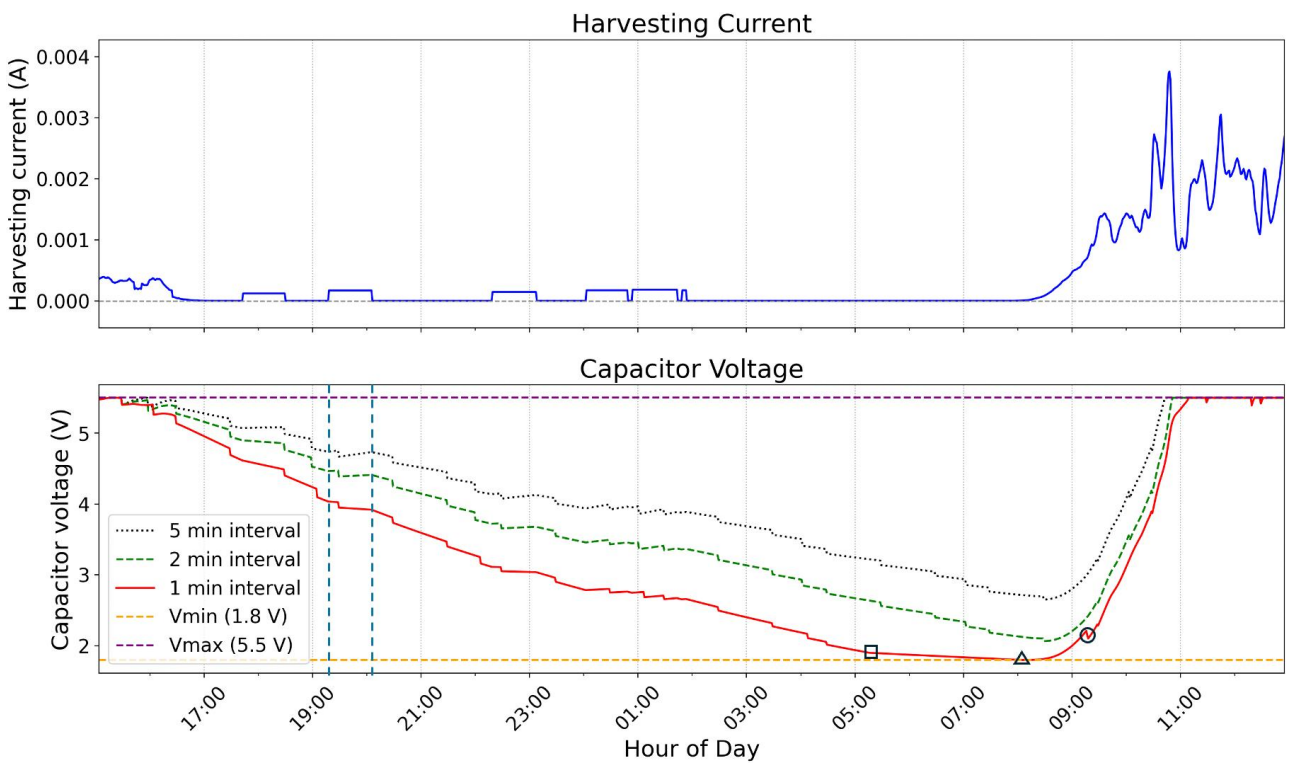

**Figure 5: Close-up analysis of the simulation (day 12 - 13) for a 2.5 F supercapacitor for different GPS fix intervals.**

*5.2.3 Comparison of GPS Fix Intervals.* Table 4 provides insights into the number of GPS fixes recorded during the simulation for the three previously discussed fix intervals. Both the 5-minute and 2-minute configurations exhibit negligible standard deviation in their daily fix counts, reflecting consistent timing and reliable scheduling. This is further supported by the absence of warm start ephemeris downloads, indicating that the system always managed to refresh its ephemeris data within the four-hour limit, avoiding fallback to warm start mode. From a performance standpoint, the 2-minute interval offers a clear advantage with 2.6 times more GPS measurements per day compared to the 5-minute interval, without sacrificing reliability. This makes it a favorable configuration when using a 2.5 F capacitor, as it enables high-resolution tracking while maintaining continuity. In contrast, while the 1-minute interval achieves even more frequent GPS updates—about 1.7 times more than the 2-minute case—this comes at a cost. The significantly higher standard deviation in daily fix counts, the occurrence of warm start ephemeris downloads, and the observed shutdown period (see Figure 5), leading to cold start, all point to system instability at this resolution. Most critically, from an application perspective, these instabilities lead to large gaps in the collected tracking data, sometimes lasting multiple hours. As a result, although the nominal fix rate is higher, the effective data coverage may be lower due to these discontinuities, posing challenges for behavioral analysis and movement pattern reconstruction in wildlife tracking.

**Table 4: GPS Fix Metrics For Different Fix Intervals**

| Metric | 5 min | 2 min | 1 min |
|---|---|---|---|
| Hot Fixes | 3517 | 9272 | 16119 |
| Hot Ephemeris | 106 | 106 | 86 |
| Warm Ephemeris | 0 | 0 | 9 |
| Cold Start | 0 | 0 | 1 |
| Total Fixes | 3623 | 9378 | 16215 |
| Average per Day | 270.46 | 700.00 | 1204.15 |
| Std Dev per Day | 0.63 | 0.00 | 70.10 |

*5.2.4 Comparison of Capacitor Sizes.* The supercapacitor size significantly affects system behavior and reliability. Figure 6 shows the simulated system behavior with different capacitor sizes while keeping the GPS fix interval constant at 2 minutes. Smaller capacitors, with their limited energy storage, show sharper voltage fluctuations and are more vulnerable to brief energy deficits. While they recharge quickly during harvesting periods, their small buffer increases the risk of short-term depletion, where the voltage drops below $V_{\text{min}}$ and causes shutdowns. This is clearly demonstrated in the 1 F case. Larger capacitors, in contrast, provide better buffering and smoother voltage dynamics, offering protection against temporary energy gaps. However, their slower charging rate makes them more susceptible to long-term depletion, where the voltage gradually declines over multiple days if daily energy intake is insufficient. While this gradual decline is not explicitly observed in our evaluation of the 5 F capacitor, we do see that it fails to fully recharge on day 4. If multiple consecutive days with poor sunlight conditions were to occur, this could lead to eventual energy exhaustion. Choosing the right capacitor size is a trade-off: small sizes favor responsiveness but risk instability, while large ones improve resilience but may struggle in low-harvest conditions. The optimal size balances short-term stability and long-term sustainability based on energy demands and environmental conditions. Based on the evaluation, both the 2.5 F and 5 F capacitors present themselves as viable options for our wildlife tracking application, supporting consistent GPS and Coulomb counter measurements at a resolution of 2 minutes while wirelessly communicating this data every hour.

# 6 CONCLUSIONS AND FUTURE WORK

This paper presented a simulation-based evaluation of an energy-neutral wildlife tracking system powered by multi-source energy harvesting, using a supercapacitor for storage. A key innovation in our design is using multiple ambient energy sources, and exploiting the dual use of the kinetic energy harvester, which contributes to the power budget and also serves as a sensor for the animal's movement intensity by sampling the harvested current. Our results show that the system can reliably collect GPS data and kinetic harvesting data every two minutes, while transmitting the aggregated results via NB-IoT every hour. Through systematic evaluation, we demonstrated how task frequency and capacitor size affect performance, energy stability, and data continuity. We found that a 2.5F capacitor offers a good trade-off between responsiveness and resilience, supporting consistent operation even under varying environmental conditions. This work is currently limited to simulation. As future work, we plan to implement and validate the system on hardware in real-world conditions. Additionally, we aim to explore adaptive scheduling strategies that better manage energy across day and night cycles, such as lowering the GPS fix rate at night or predicting low-harvest periods to conserve energy. Finally, the system can be evaluated for other tracking use-cases, since it is designed to be highly configurable and modular.

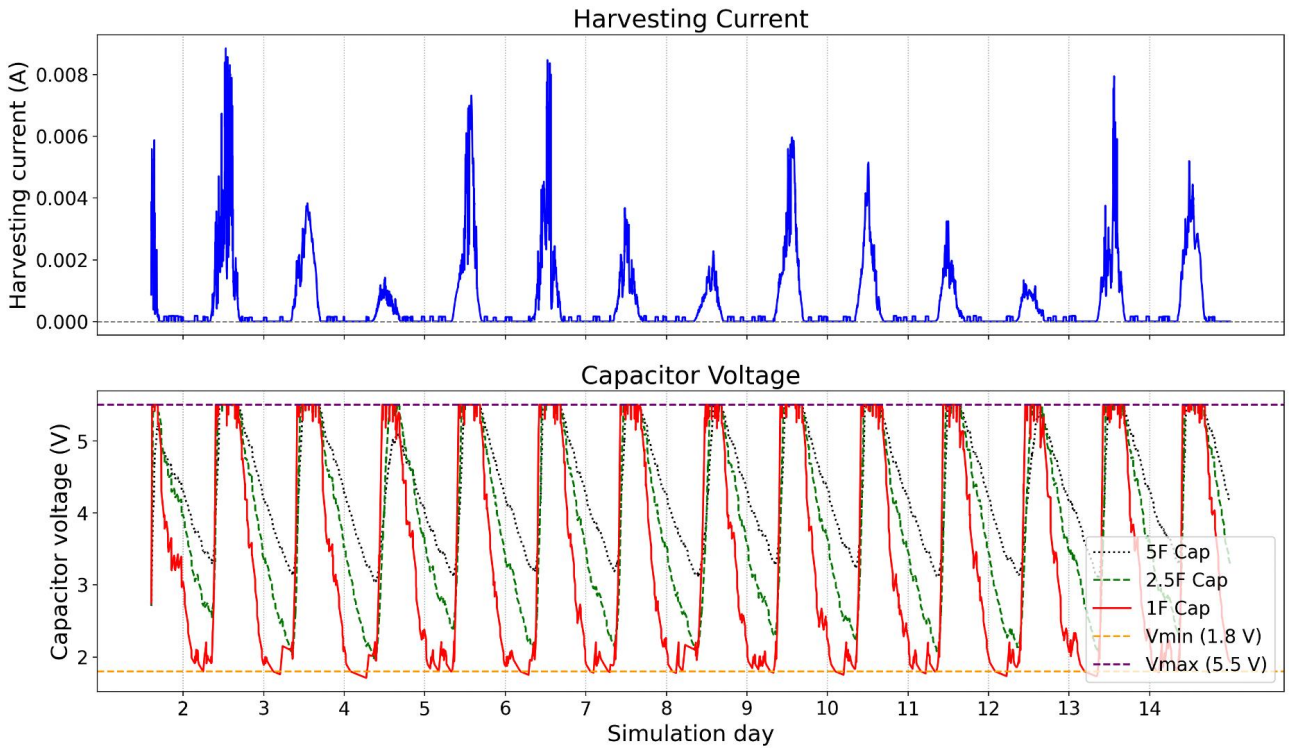

**Figure 6: Full simulation overview of the combined harvesting current and the capacitor voltage for three different capacitor sizes with a GPS fix interval of 2 minutes.**

## ACKNOWLEDGEMENTS

This research was funded by the IoBaLeT and AMBIENT-6G projects. IoBaLeT (Sustainable Internet of Batteryless Things) received funding from the Fund of Scientific Research Flanders (FWO) under grant agreement number S001521N. The AMBIENT-6G project has received funding from the Smart Networks and Services Joint Undertaking (SNS JU) of the European Union's Horizon Europe research and innovation programme under Grant Agreement No 101192113.